\documentclass[%
 reprint,
superscriptaddress,
 amsmath,amssymb,
 aps,
pra,
]{revtex4-2}

\usepackage{graphicx}% Include figure files
\usepackage{dcolumn}% Align table columns on decimal point
\usepackage{bm}% bold math
\usepackage{placeins}
\usepackage{float}
\usepackage{svg}
\usepackage{afterpage}
\usepackage{upgreek}
\usepackage{physics}

\begin{document}

\begin{abstract}
Superconducting quantum computers leverage superconducting qubits, systems with well-defined energy levels, the spacing of which is determined by the properties of embedded Josephson junctions. Indeterminacy of Josephson junction manufacturing limits large scale superconducting circuits, however junction-by-junction tuning affords the possibility of an order of magnitude improvement of determinacy - although the origin of this tuning effect is uncertain.
Understanding how tuning techniques may physically modify the tunnel barrier of a Josephson junction is important and will enable these techniques to be optimised.
We develop a model of junction tuning based on depinning theory to interpret a phase diagram of tuning rate. We extract the dependence on temperature, time-varying voltages and oscillation frequency. 
Comparing the effect of junction tuning on room-temperature IVs, electrical breakdown and frequency of higher energy levels of transmon qubits indicates that the tuning process leaves the barrier thickness unchanged but modifies its composition.

\end{abstract}

% \preprint{APS/123-QED}

\title{A Depinning Model for Josephson Junction Tuning}% Force line breaks with \\

\author{Oscar~W.~Kennedy}\email{okennedy@oqc.tech}
\affiliation{%
 Oxford Quantum Circuits, Thames Valley Science Park, Shinfield, Reading, United Kingdom, RG2 9LH}%
 
\author{Jared~H.~Cole}\email{jared.cole@rmit.edu.au}
\affiliation{Theoretical, Computational, and Quantum Physics, School of Science, RMIT University, Melbourne, Australia}
 
\author{Connor~D.~Shelly}\email{cshelly@oqc.tech}
\affiliation{%
 Oxford Quantum Circuits, Thames Valley Science Park, Shinfield, Reading, United Kingdom, RG2 9LH}%

\date{\today}% It is always \today, today,

\maketitle

%\tableofcontents

Quantum computers with high qubit count, low error rates and error correcting protocols promise new computational capabilities. Processors comprised of superconducting qubits are a leading platform for utility scale quantum computation~\cite{arute2019quantum, chow2021ibm, google2023suppressing} but building these at scale and with high quality poses many challenges.  
One component which poses manufacturing difficulties is the Josephson junction (JJ). These are most often based on stacks of Al/AlO$_{\rm x}$/Al and are typically made by shadow angle evaporation~\cite{kreikebaum2020improving} where a bottom layer of thin-film aluminium is oxidised before being coated by a second layer of aluminium. 
%Typically JJs are characterised by a room temperature resistance which is indicative of their cryogenic properties but more readily measured~\cite{kreikebaum2020improving} and state-of-the-art resistance variation sits at $\sim$2-3\%~\cite{osman2023mitigation, acharya2024integration, van2024advanced}.
% Uncontrolled variation in as-fabricated JJs has been identified as a bottle-neck in the manufacture of quantum processors which match design specifications~\cite{hertzberg2021laser, wang2024precision, muthusubramanian2024wafer, osman2023mitigation, zhang2022high, morvan2022optimizing}.
State-of-the-art variation in as-fabricated JJs currently limits junction resistance accuracy to $\gtrapprox $2~\%~\cite{hertzberg2021laser, wang2024precision, muthusubramanian2024wafer, osman2023mitigation, zhang2022high, morvan2022optimizing}. The problem of junction variation will likely be exacerbated in larger wafers necessitated by larger processors and by the use of long-range couplers~\cite{du2025hardware}. Simulations show that current manufacturing accuracy is insufficient to build large, high quality, quantum processors~\cite{morvan2022optimizing, osman2023mitigation, hertzberg2021laser}. Improvements to frequency assignment enabled by tuning protocols, have been shown to improve processor yield experimentally~\cite{zhang2022high}.
Not only are junctions challenging to make with tight tolerance, but their properties change over time after their manufacture in a process known as aging~\cite{koppinen2007complete} which complicates processor design. 
In order to address this variation a family of techniques which allow post-manufacture parameter adjustment have been developed including modifying the junction by laser irradiation~\cite{hertzberg2021laser, kim2022effects}, electron beam irradiation~\cite{balaji2024electron} and alternating bias assisted annealing (ABAA)~\cite{pappas2024alternating}. These techniques, referred to here as tuning, cause the resistance of junctions to increase only. 

In this letter, we explore the voltage-temperature-frequency space, facilitated by a simplified lock-in tuning technique adapted from~\cite{pappas2024alternating}. We interpret the resultant phase diagram using depinning theory which applies to a wide range of systems in which there is an interplay between disorder, temperature and applied stresses. This is seen in domain walls in magnets, the movement of vortices in superconductors, charge movement in Josephson junction arrays, dislocation movement, crack formation and even the formation of avalanches in sand piles~\cite{Wiese:2022, Brazovskii:2004,Vogt:2014,Vogt:2015,Cedergren:2017}. We find good agreement between our depinning model and the three-dimensional phase diagram. 
We compare tuned and untuned junctions measuring tunneling IV curves, breakdown voltages and Josephson harmonics~\cite{willsch2024observation}. 

\begin{figure}
    \centering
    \includegraphics[width=\linewidth]{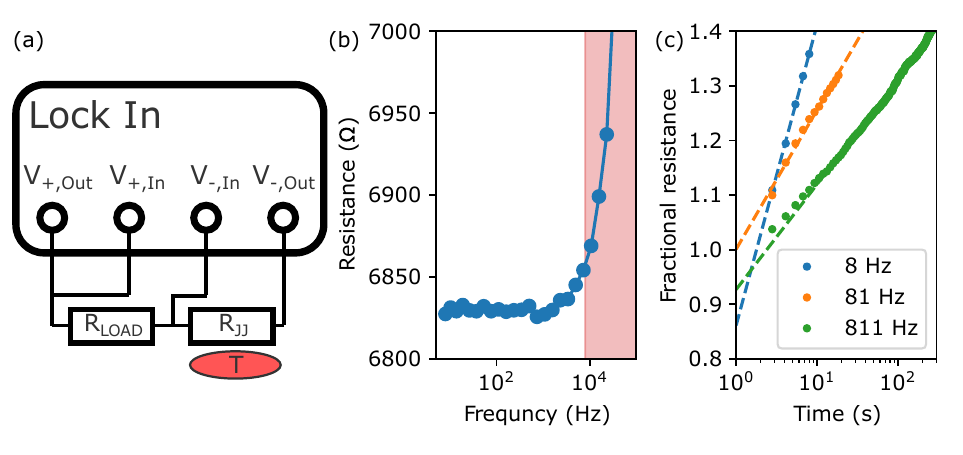}
    \caption{(a) A schematic of the experimental setup showing a lock-in amplifier applying a voltage across a load resistor and a JJ in series. A heater controls the temperature of the JJ. (b) Resistance as a function of the frequency of the lock-in amplifier. Above $\sim$8~kHz the nominal resistance of the junction increases rapidly with frequency, a region we omit from study. (c) Three example tuning curves where the fractional resistance of the JJ (i.e. resistance normalised to that at the start of the protocol) against the elapsed time in the protocol performed at 80~$^\circ$C and a tuning voltage of 0.95~V. Dashed lines are a fit of the data to Eq.~\ref{eq:tuning}.}
    \label{fig:freq}
\end{figure}

We build a setup shown schematically in Fig.~\ref{fig:freq}~(a) where we connect a JJ in series with a load resistor, creating a potential divider, and connect this circuit between the output ports of a Zurich Instruments MFLI lock-in amplifier. Measuring the voltage drop across either of these elements allows us to infer the resistance of the JJ, the property we optimise in this process. We can control both the amplitude and the frequency of the AC voltage output from the lock-in amplifier. The sample containing JJs is placed on a Peltier heating stage to control the temperature of the JJ during this process. JJs are fabricated in Ref~\cite{acharya2024integration}, Dolan bridge junctions with an area of $\sim$84,000~nm$^2$ are fabricated by shadow evaporations before and after a static oxidation step. An atomic force micrograph of a typical junction is shown in the supplemental materials~\cite{supp} which includes Ref.~\cite{dobrovinskaya2009properties}.

We measure the resistance of the JJ as a function of the lock-in frequency (Fig.~\ref{fig:freq}~b) and find that we measure a constant value up to a frequency of $\sim$8~kHz, when the measured resistance increases. This increase in resistance is a measurement artefact due to RC filtering based on a computed cut-off frequency of $\sim$40~kHz, with most capacitance arising from coaxial cabling - a series resistance of 15~k$\Omega$ (10~k$\Omega$ load and $\sim$5~k$\Omega$ junction resistance)  together with 3~m of BNC cables (typical capacitance of $\sim$80~pF/m) gives an RC frequency of 44 kHz. We confine our frequency operating window to frequencies of 1~kHz and below to ensure we infer the correct resistances of the JJs. 

In Fig.~\ref{fig:freq}~(c) we show how resistance changes during the tuning protocol for three frequencies; 8~Hz, 81~Hz and 811~Hz. We fit these tuning curves phenomenologically to a model where the resistance increases logarithmically with time
\begin{equation}
    R(t)/R_0 = a \log(ct) 
    \label{eq:tuning}
\end{equation}
where $R_0$ is the resistance at the start of the protocol and both $a$ and $c$ are fit parameters, with $a$ relating to the speed of tuning. We find this model fits all measured tuning curves well, and that tuning increases in speed with decreasing frequency. 
On some occasions, not shown in Fig.~\ref{fig:freq}, the junction fails during the tuning process, becoming a short circuit. 
The rate of junction failures for a given process is another parameter we explore. We show that the behaviour (speed and failure rate) of junctions undergoing voltage tuning can be understood qualitatively in terms of depinning theory.

The classic picture of depinning theory as applied to disordered media invokes two phases, a pinned static phase where the system is in a relaxed local minimum and a sliding phase where local regions or domains within the system have a nonzero average velocity~\cite{Middleton:1991, Nattermann:2001}. The domains of depinning theory are a general concept, which may sometimes be instantiated by magnetic domains specifically, but this commonality in terminology is not fundamental. For example, domains may similarly be regions of sand grains, or any other type of effective degree of freedom that `depins' in a disordered system.
In the pinned phase, any static or oscillating field can induce local microscopic changes, which can act as a dissipation channel~\cite{muller2019towards}, but not large scale movement or restructure. 
Above some threshold applied field (the `depinning threshold') the movement of these local domains has some finite average velocity and can ultimately lead to avalanche behaviour.

We consider the junction dielectric to host domains which are trapped local minima. 
These domains may be 0D vacancies/interstitials, 1D chains of such vacancies, or 2D boundary regions of lower density amorphous material or something else - the depinning treatment here is agnostic to the underlying microphysics. 
Applying sufficient field or temperature can move these domains, surmounting some effective potential barrier. Following Nattermann \emph{et al.}~\cite{Nattermann:2001}, we define the depinning temperature $T_{\rm P}$ where the thermal energy is equivalent to the barrier height. The depinning voltage $V_{\rm P}$ is the point at which there is sufficient applied field for domains to continuously move through the material, resulting in an avalanche.

\begin{figure*}
    \centering
    \includegraphics[width=\linewidth]{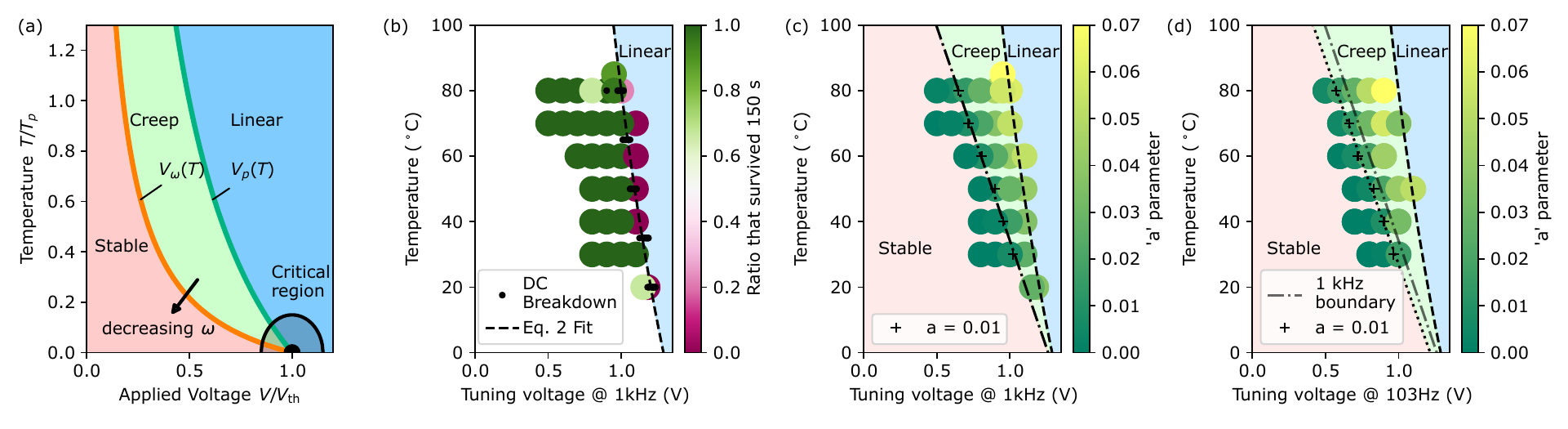}
    \caption{(a) Generic phase diagram for depinning theory, showing three operating regimes, where the creep-regime is where tuning occurs. (b) Threshold voltage phase diagram as a function of tuning voltage and temperature. Measurements of the fraction of junctions failing during processing show this boundary. Overlaid are DC measurements of breakdown which follow the same boundary separating the creep and linear regimes. These data are fit by Eq.~\ref{eq:vp} (c) 1~kHz and (d) 103~Hz resistance tuning speed phase diagram. The parameter $a$ from Eq.~\ref{eq:tuning} is shown using a colour scale. The boundary between the creep and linear regime found by fitting DC breakdown in (b) is shown in both phase diagrams, as is a straight line fit to points where the interpolated value of a is 0.01 showing the stable/creep boundary. In (d) we show the boundary for both 1000~Hz and 103~Hz showing that with decreasing frequency the boundary has moved to lower voltages and temperatures.}
    \label{fig:depinning}
\end{figure*}

The regions of interest in a generic depinning theory are illustrated in Fig.~\ref{fig:depinning}~(a). The linear response or `running' region (blue shaded area) applies for constant fields greater than the temperature dependent depinning voltage $V_{\rm P}(T)$. Any constant applied field less than this will not result in a significant change in the structure of the material, it will be `pinned' (the red shaded area). The zero temperature limit of $V_{\rm P}(T\rightarrow 0)$ is the threshold voltage $V_{\rm{th}}$.

The general depinning threshold ($V_{\rm P}(T)$) which separates these two regions can be defined for temperature $T$ and voltage $V$ by the relation
\begin{equation}
    T = T_{\rm P}\frac{V_{\rm{th}}}{V} \left(1-\frac{V}{V_{\rm{th}}} \right)^{1/\mu} 
    \label{eq:vp}
\end{equation}
where $\mu$ is a constant which depends on the effective dimensionality of the elastic object (for 1-2 dimensional domains in a 3D medium, it can be approximated to 1)~\cite{Nattermann:2001, Nattermann:2004}.

When an oscillatory field is applied, a 3rd regime is formed in the phase diagram. The so-called `creep' region (green region of Fig.~\ref{fig:depinning}~a) is the region in which domains can most easily move under application of an oscillatory field. 
In the creep region, the oscillating field results in hysteretic behaviour~\cite{Schuetze:2011, Brazovskii:2004}. 

A general expression for the creep threshold $V_\omega(T)$ is difficult to derive without an explicit microscopic model. Nattermann~\textit{et al.} suggested~\cite{Nattermann:2001, Nattermann:2004} this boundary can be found by modifying Eq.~\ref{eq:vp}, replacing $T_{\rm P} \rightarrow T_{\rm P}/\Lambda$ in the adiabatic limit. In this case $\Lambda=\ln(1/\omega_0\tau)$ depends on the oscillation frequency $\omega_0$ and the characteristic timescale $\tau$ required for a domain to overcome the barrier to move from its local equilibrium. In what follows, we find this significantly overestimates the shift due to variation in $\omega_0$ for sensible estimates of $\tau$, suggesting a more rigorous treatment is required that goes beyond the adiabatic approximation.

Interpreting the junction tuning results in terms of depinning suggests a) if $\omega_0 \tau < 1$ the creep region is larger for lower frequencies, and b) in the creep regime, there is a linear movement of defects without leading to full avalanche. This explains why lower frequencies result in greater resistance shifts for fixed values of temperature and voltage amplitude. At some low-frequency-limit we anticipate this trend stops, as domain movement saturates before the field polarity is reversed, presenting a new energy landscape to the mobile domains.
We can also understand the junction breakdown in terms of the depinning avalanche process for $V>V_{\rm P}(T)$, which is both temperature and local microstate dependent.

To explore our interpretation we tune junctions with varying voltage and temperature at frequencies of both 1~kHz and 103~Hz. For each parameter set we tune at least three junctions.  We fit each of these tuning curves to Eq.~\ref{eq:tuning} and extract the speed parameter $a$. We also compute how many junctions fail prior to 150~s of tuning process. We present this data as a function of temperature and voltage amplitude, Fig.~\ref{fig:depinning}~(b). A clear boundary is observed at a tuning voltage of $\gtrsim 1$~V. This boundary is matched by DC measurements of breakdown voltage performed at varying temperature. We therefore associate this boundary with the depinning threshold $V_{\rm P}(T)$ and fit the DC breakdown to Eq.~\ref{eq:vp}.

We continue to map out the phase diagram by measuring the tuning `speed' parameter, $a$ in Eq.~\ref{eq:tuning}. Operating at 1~kHz, we vary the temperature and voltage amplitude of the tuning process. In Fig.~\ref{fig:depinning}~(c) we plot $a$, finding a similar boundary, now at a lower voltage, consistent with the expected behaviour of $V_\omega(T)$. This process is then repeated at a different tuning frequency (103~Hz) with data shown in Fig.~\ref{fig:depinning}~(d), showing the expected increase in the creep regime with reducing frequency.

To fit the boundaries between regimes quantitatively is more difficult, as it requires estimates for $\tau$, $\mu$, $T_{\rm P}$ and $V_{\rm{th}}$ and a better understanding of the relationship between depinning velocity and the tuning speed parameter. We can estimate $V_{\rm{th}}$ from junction breakdown measurements at a given temperature. 
Some indications for the depinning temperature can be given by observations of where the rate of thermal aging starts to increase.
To give an indication of likely temperature scales, in ~\cite{Imamura:1992,Oliva:1994,Migacz:2003} changes are first seen $>125~^\circ$C and large scale junction modification is observed at $>400~^\circ$C in~\cite{Pop:2012,Julin:2010,Koppinen:2007, korshakov2024aluminum}. 

Using the approach~\cite{Nattermann:2001, Nattermann:2004} given in Eq.~\ref{eq:vp}, along with a finite value of $\Lambda$, we compare to the stable/creep boundaries shown in Fig.~\ref{fig:depinning}c) and d). We find that while the direction of the experimental shift with bias frequency is expected, its magnitude is actually much smaller than depinning theory would suggest (or corresponds to unrealistically small $\tau$ -- see supplementary material~\cite{supp}). This suggests that the dynamics in the creep regime are more complicated than in previous studies of depinning.

Fig.~\ref{fig:depinning}~(a) indicates sharp lines demarking regions of the phase diagram, however, at this stage it is not clear that these boundaries are true phase transitions. The speed parameter $a$ increases smoothly with increasing tuning voltage (the data is shown with alternate visualisation in the supplementary materials~\cite{supp}). This may be indicative of a second order phase transition between the stable and creep regimes.

Mapping out this phase diagram we have identified room temperature working points with appreciable process speed as in Ref.~\cite{wang2024precision}. This offers multiple advantages, key among them is that junctions which are not tuned undergo no heating and associated aging, a significant benefit for processor design.
The exact mechanism of the tuning, as well as the nature of the domains being depinned, is currently unknown. More detailed mapping of the phase diagram, combined with the statistics of many junctions, will provide invaluable data to compare to potential models of junction tuning.

While the depinning picture provides a qualitative description of the observed dynamics, a quantitative model is needed to understanding the underlying microscopic processes.
Time-dependent dielectric breakdown has been extensively studied in silica-based dielectrics~\cite{mcpherson2012time, moxim2022atomic, mcpherson2003thermochemical}, in magnetic tunnel junctions~\cite{amara2012modelling} and low-k dielectrics~\cite{borja2016dielectric, borja2012impact}, as well as AlOx films~\cite{schaefer2011dielectric, kolodzey2000electrical, Tolpygo:2008, de1976dielectric}. In these materials, a variety of breakdown mechanisms have been studied, including charge trapping within the barrier, time modulation of this charge and electric field-induced distortion of bonds in the barrier. In particular studies of copper ion migration~\cite{borja2016dielectric, borja2012impact} have considered the influence of 
both convective and diffusive flux of ions. These correspond to diffusion driven by the electric field itself, or by the stoichiometry gradient respectively. This suggests a possible two-process mechanism for breakdown and tuning of Josephson junctions. Namely that breakdown is enabled by the convective movement of atoms from within the barrier in the direction of the applied field, while the tuning of the resistance at lower fields is related to the redistribution of oxygen or defects from within the barrier via diffusion - mediated by the oscillating field. To test this possibility, we investigate the average barrier characteristics before and after tuning.

\begin{figure}
    \centering
    \includegraphics[width=\linewidth]{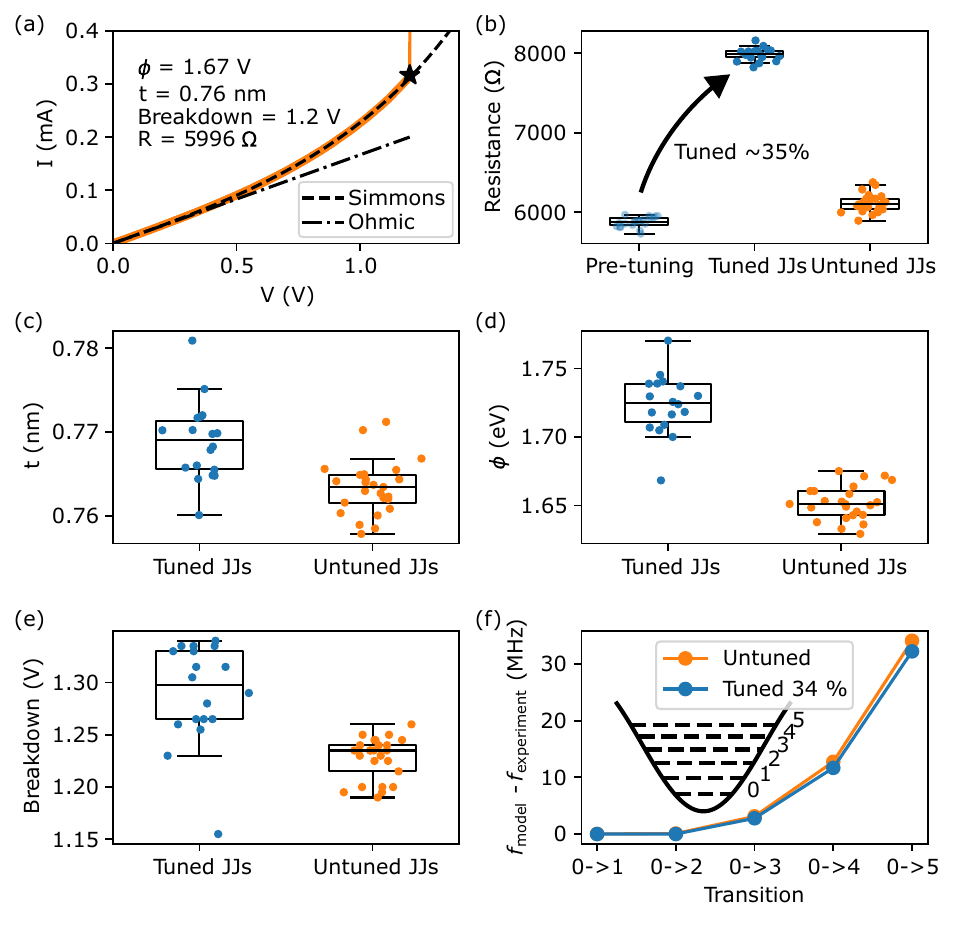}
    \caption{Investigation of the effect of tuning on DC IVs, electric breakdown and Josephson harmonics. (a) An example IV from an untuned JJ where voltage is ramped up from 0~V. It is fit to the Simmons model to determine model nominal barrier thickness and height. It is also fit to Ohm's law to determine junction resistance. Breakdown voltage is extracted where the current suddenly jumps and is indicated with a star. (b) The resistances of the junctions which have been tuned vs. those that have only been subject to the thermal history associated with tuning. The junctions are tuned an average of $\sim$35\% with the resistance prior to tuning shown in transparent blue markers. (c) The thickness extracted from the Simmons model, (d) the barrier height extracted from the Simmons model and (e) the breakdown voltage are all shown comparing tuned and untuned junctions. (f) The results from Josephson harmonics experiments performed on a tuned and an untuned qudit. The difference between the frequencies predicted by the transmon Hamiltonian and those found experimentally are plot for the different transitions. }
    \label{fig:breakdown}
\end{figure}

Following our recent work~\cite{kennedy2025analysis} we measure the electrical breakdown of junctions by ramping a DC voltage across these junctions from 0 to 1.7~V recording the current that flows through each junction as a function of voltage (Fig.~\ref{fig:breakdown}~a). We fit the linear portion of this IV to extract the resistance of the JJ and fit the full IV to the Simmons model~\cite{simmons1963generalized} fixing the area to that found by AFM measurements 8.86$\times10^4$~nm$^2$. At some voltage the junction fails and becomes short circuit which is recorded as the breakdown voltage (black star in Fig.~\ref{fig:breakdown}~a). 
We compare two sets of junctions, distinguished by their tuning, with resistances shown in Fig.~\ref{fig:breakdown}~(b). The first set of 20 junctions is tuned at 85~$^\circ$C with a voltage of 0.92~V targeting 30~\% total tuning and find an average tuning of 34.6 $\pm$ 0.8~\%. This systematic overshoot occurs due to continued resistance changes over time, after we have hit our target value, as shown in Ref.~\citenum{wang2024precision}. During this process two junctions failed. The second set of 25 junctions are fabricated on the same chip, undergo the same 85~$^\circ$C heating as the tuned junctions, but are not tuned by an alternating voltage. Despite undergoing no electrical tuning they are $\sim$4~\% higher in resistance than the junctions before tuning due to the thermal processing they undergo. This shows why a room temperature process (as shown in the supplementary materials~\cite{supp}) is advantageous as heating chips above storage temperatures causes uncontrolled changes in junction properties alongside the intentional tuning process. 

In Fig.~\ref{fig:breakdown}~(c) we see a small increase to the average fitted barrier thickness with distributions mostly overlapping for tuned and untuned junctions. In Fig.~\ref{fig:breakdown}~(d) we see a clear increase in the barrier height for the junctions which have been tuned relative to those that have not. These results indicate that the tuning process modifies the atomic structure of the barrier more substantially than its thickness. 

In Fig.~\ref{fig:breakdown}~(e) we show that on average junctions which have undergone tuning have a larger breakdown voltage, increasing by $\sim$5~\%. However, despite the increase in average breakdown voltage, the lowest breakdown voltage recorded was from a junction that had undergone tuning. The breakdown voltage is given by 
\begin{equation}
    V_{\rm BD} = \rm{min}\{t(x, y) E_{\rm BD}(x, y)\}
\end{equation}
where $t(x, y)$ is the local thickness of the barrier and $E_{\rm BD}(x, y)$ is the local dielectric strength of the barrier. Both of these values may, in principle, vary across the barrier~\cite{Cyster:2020,bayros2024influence} and be modified by tuning. 
% In our recent work~\cite{kennedy2025analysis} we show that by assuming constant $E_{\rm DS}$, we can model the standard deviation of breakdown voltages by assuming it occurs at the thinnest points in the barrier. 
% Given typical barrier thicknesses of $\sim$2~nm, a $\sim$5~\% change in thickness represents $\sim$0.1~nm, less than the typical ionic diameter in these solids~\cite{dobrovinskaya2009properties}. 
In the depinning model we describe above domains are mobile under the application of electric fields. These fields will be the strongest at the thinnest regions of the barrier which means that the the creep rate will also be greatest at these thinnest points.
Such weak points in the barrier have been suggested to contribute to non-idealities~\cite{bayros2024influence, willsch2024observation, Greibe:2011} and breakdown~\cite{Tolpygo:2008} of the JJs. An approach that reduces the non-idealities associated to pin holes or weak points may have advantages in homogenising the current-phase relations of junctions. 

The higher-energy states of transmon qubits have recently been suggested to be sensitive to the nanostructure of the barrier in JJs~\cite{willsch2024observation}. 
In Fig.~\ref{fig:breakdown}~(f) we show the results of measuring two coaxmon qubits~\cite{rahamim2017double} where one qubit has had a junction tuned by 34~\% and the other has experienced the same thermal treatment but not been tuned. The results of Ref.~\citenum{willsch2024observation} would suggest that by substantially modifying the barrier we might expect to see some difference to the frequencies of these higher levels. 
We follow the protocol from Ref.~\citenum{wang2024systematic}, treating the coaxmon as a qudit and tuning up a series of gates across the 6 energy levels shown schematically in Fig.~\ref{fig:breakdown}~(f). We compute the Josephson energy and charging energy using qutip~\cite{johansson2012qutip} by considering the first two transition frequencies and constraining them to match the results from the transmon Hamiltonian
\begin{equation}
    H = 4 E_{\rm C} \hat{n}^2 - E_{\rm J}\cos{\hat{\phi}}
\end{equation}
where $\hat{n}$ and $\hat{\phi}$ are the charge and phase operators respectively and $E_{\rm C}$ and $E_{\rm J}$ are the charging and Josephson energies~\cite{koch2007charge}. The tuned (untuned) coaxmon have  junction resistances of 5802~$\Omega$ (6021~$\Omega$), charging energies of 166~MHz (168~MHz) and Josephson energies of 23.2~GHz (22.2~GHz). 
In Fig.~\ref{fig:breakdown}~(f) we show the difference between the measured transition frequencies and the computed transition frequencies for the different qudit states for both a tuned and untuned coaxmon. We find the relative difference in model/experiment discrepancies for tuned and untuned junctions to be small relative to lab-to-lab variation or even qubit-to-qubit variation from a single institute presented in Ref.~\cite{willsch2024observation} and therefore conclude that the tuning process has not substantially modified the Josephson harmonics in this instance. Statistical studies comparing multiple tuned and untuned qubits may allow small effects to be resolved in future studies.

By mapping out the voltage-temperature-frequency phase diagram of tuning under oscillatory bias we have shown that the tuning process is qualitatively described by depinning theory. The domains within the amorphous oxide barrier creep between local minima driven by both temperature and the oscillating field. Depinning theory by itself does not point to a particular origin of these local domains, as it is consistent with small scale defects~\cite{Strand:2024}, movement of grain boundaries~\cite{Oh:2025} or other microscopic models~\cite{muller2019towards}.
Nonetheless by interpreting the results in terms of depinning theory, the working point can be chosen to improve the junction tuning protocol, choosing how deeply into the creep-region one wishes to operate.
We find a set of parameters that allow junction tuning at room temperature with appreciable speed, where untuned junctions are also not subject to heating and associated modifications. 

We show that tuning modifies, not only the resistance of the junctions, but also the barrier height in the junction and the breakdown voltage of the junctions.  
We contrast this to a measurement of `Josephson Harmonics' which appear unchanged comparing tuned and untuned junctions. 
The interpretation in Ref.~\citenum{willsch2024observation} suggests that deviations in these harmonics can be ascribed to the thinnest points of junctions. 
Taken together our measurements suggest that the depinning process modifying the junction resistance is not simply `filling in' the thinnest points of barriers, but fundamentally modifies the junction microstructure.
%Conversely, our measurements therefore suggest that the thinnest points of barriers remain unchanged upon tuning the junction a substantial fraction. 
% Reconciling these observations will allow a deeper understanding of what constitutes domains in these junctions. 

\begin{acknowledgments}
We extend our thanks to the OQC Fabrication team for discussions relating to this work, the fabrication of junctions and AFM measurements of junction area. We acknowledge the OQC Hardware team for developments across the stack which enabled these measurements.
The authors thank Dave Pappas and Simon Brown for useful discussions.
The authors thank Peter Leek and Jonathan Burnett for reviewing this manuscript. 
Computational resources were provided by the Australian National Computational Infrastructure facility (NCI).
We thank the Royal Holloway University of London SuperFab Facility for their support.

\end{acknowledgments}

\bibliography{bibliography}% Produces the bibliography via BibTeX.

\newpage

\appendix
\end{document}

% \preprint{APS/123-QED}

\title{Supplemental Materials for A Depinning Model for Josephson Junction Tuning}% Force line breaks with \\

\author{Oscar~W.~Kennedy}\email{okennedy@oqc.tech}
\affiliation{%
 Oxford Quantum Circuits, Thames Valley Science Park, Shinfield, Reading, United Kingdom, RG2 9LH}%
 
\author{Jared~H.~Cole}\email{jared.cole@rmit.edu.au}
\affiliation{Theoretical, Computational, and Quantum Physics, School of Science, RMIT University, Melbourne, Australia}
 
\author{Connor~D.~Shelly}\email{cshelly@oqc.tech}
\affiliation{%
 Oxford Quantum Circuits, Thames Valley Science Park, Shinfield, Reading, United Kingdom, RG2 9LH}%

\date{\today}% It is always \today, today,

\maketitle

%\tableofcontents

\section{Josephson Junction Geometry}
\begin{figure}
    \centering
    \includegraphics[width=\linewidth]{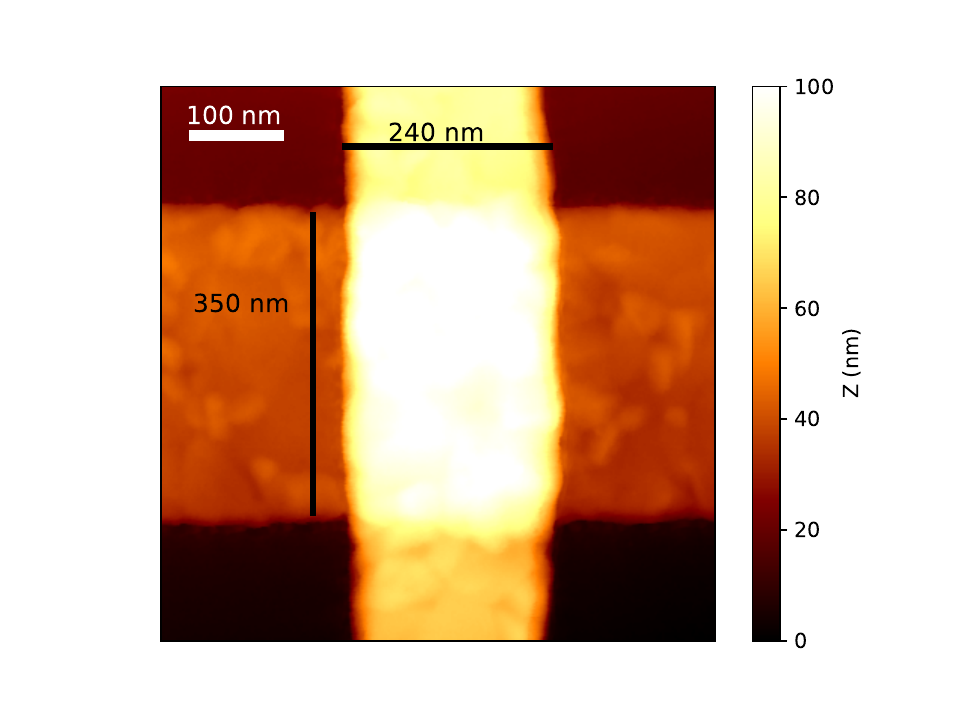}
    \caption{AFM of a typical junction used in these experiments. The junction is a Dolan bridge type junction with $\sim$240$\times$350~nm lateral dimensions. }
    \label{fig:afm_junction}
\end{figure}

In Fig.~\ref{fig:afm_junction} we show an atomic force microscope (AFM) micrograph of a typical junction fabricated on a sapphire substrate used in these tuning experiments. Junctions are formed in the Dolan bridge geometry. A bottom lead is deposited, oxidised in situ with a static oxidation process before the top aluminium lead is deposited. Dimensions are indicated on the AFM image. Junctions are deposited in 5x10 arrays with identical target dimensions prior to tuning. 

\section{Room temperature tuning}
\begin{figure}
    \centering
    \includegraphics[width=\linewidth]{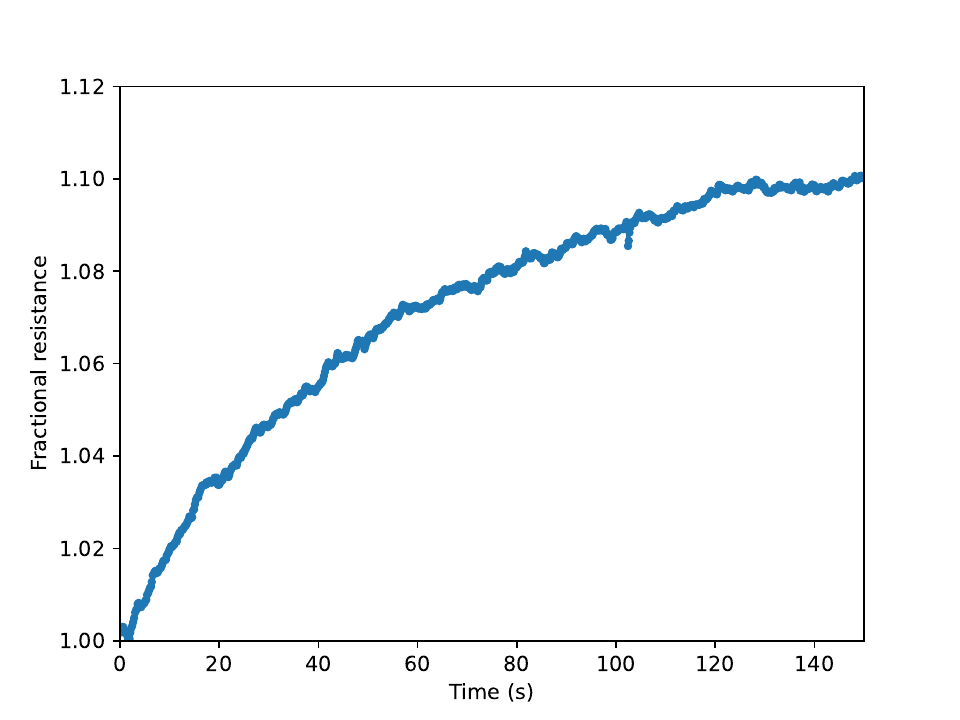}
    \caption{Fractional resistance vs. time of a junction tuned at room temperature using a tuning voltage of 1.15~V. We show a $\sim$10\% tuning range in O(100 s). }
    \label{fig:rt_tune}
\end{figure}

In Fig.~\ref{fig:rt_tune} we show a single junction being tuned at room temperature. Here we show tuning of an appreciable fraction of the junction resistance, greater than the resistance variation common in as-fabricated junctions. These processes could therefore be used to correct resistance variations in as-fabricated junctions. The maximum tuning range and tuning speed could be further optimised in future studies. 

\section{Power law fitting}
\begin{figure}
    \centering
    \includegraphics[width=\linewidth]{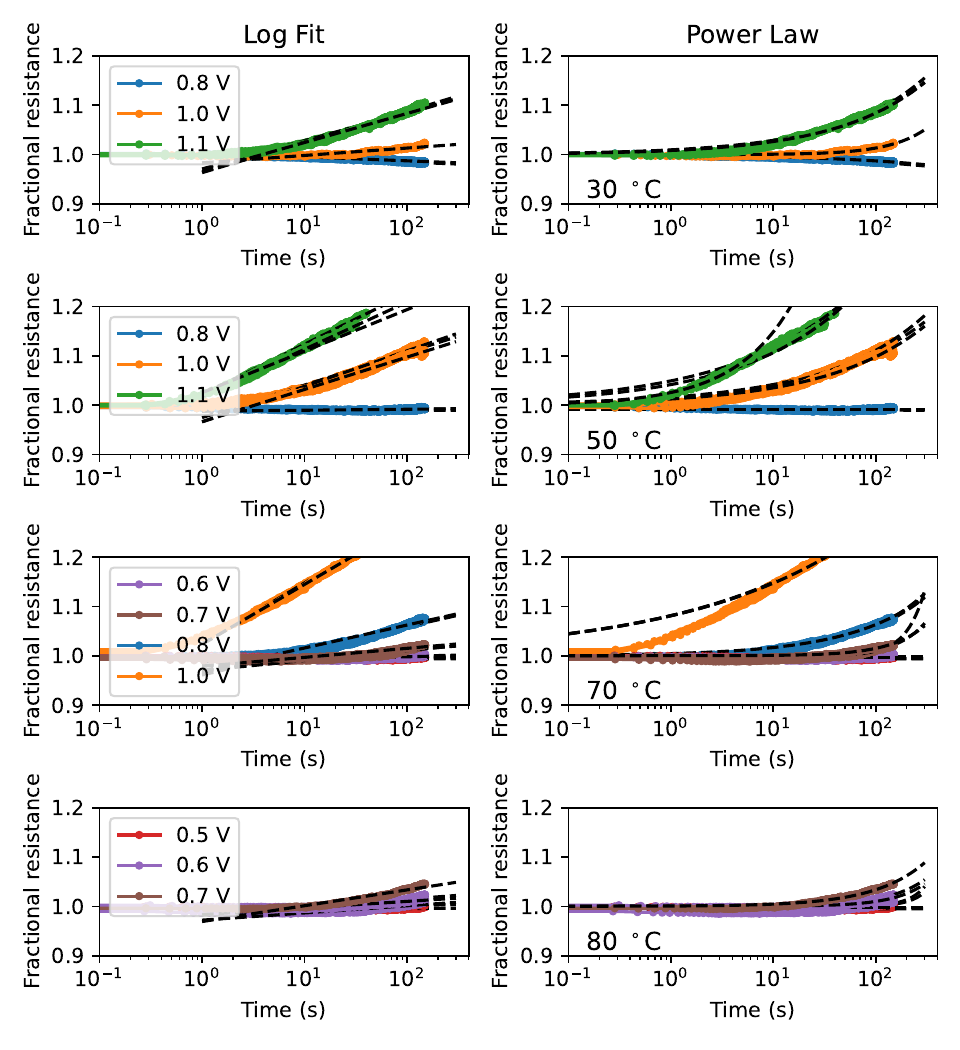}
    \caption{Comparison between fits to logarithmic function and power law functions for tuning behaviour. The y axis of each panel is the normalised resistance of the junction and the x axis is time on a logarithmic axis. Black dashed lines show the logarithmic and power law fits which fit quite well to most the data (excluding very short times).}
    \label{fig:power_law}
\end{figure}

We investigate fitting the tuning curves to both the phenomenological logarithmic tuning curve $R(t)/R_0 = a \log(ct)$ and also fitting the tuning to a power law $R(t)/R_0 = a t^n + 1 $. We present the results of the two fitting protocols in Fig.~\ref{fig:power_law}. Both functions fit the data relatively well but there are fewer fits which appear to fail for a logarthimic function. In a power law, it is less obvious how to interpret the different fit parameters to a single speed function and hence we use the logarithmic function in our analysis. 

\section{Phase diagram data}
\begin{figure}
    \centering
    \includegraphics[width=\linewidth]{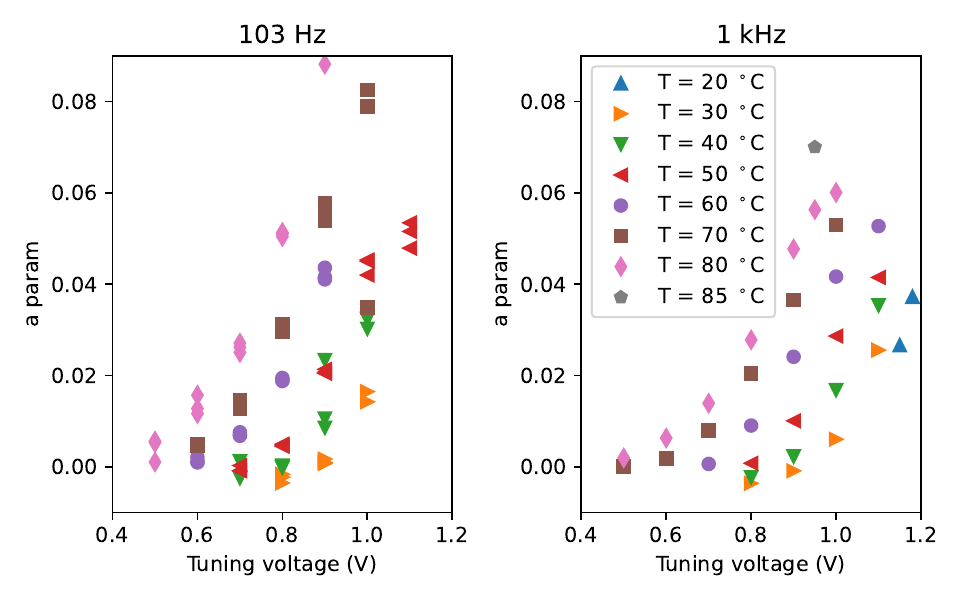}
    \caption{The experimental tuning speed data (or a parameter) from the 2D phase diagrams shown in Fig.~2 main text is presented here so that the reader can clearly see trends in this value. It is plot as a function of tuning voltage with different temperatures indicated by different marker colours }
    \label{fig:phase_diag_rawdata}
\end{figure}
We show the data from the phase diagrams presented in Fig.~2 main text, as markers in Fig:~\ref{fig:phase_diag_rawdata}. We can see that the speed parameter `a' smoothly increases from 0, without showing marked discontinuities. For some process parameters we observe an apparent negative speed, i.e. that resistance drops over time. This is not something that is sustained across the end of the tuning process and junctions always recover to, and typically over-shoot, the junction resistance at the start of this process. The origin of this process is not immediately clear. 

\section{Self heating during tuning}
\begin{figure*}
    \centering
    \includegraphics[width=0.75\linewidth]{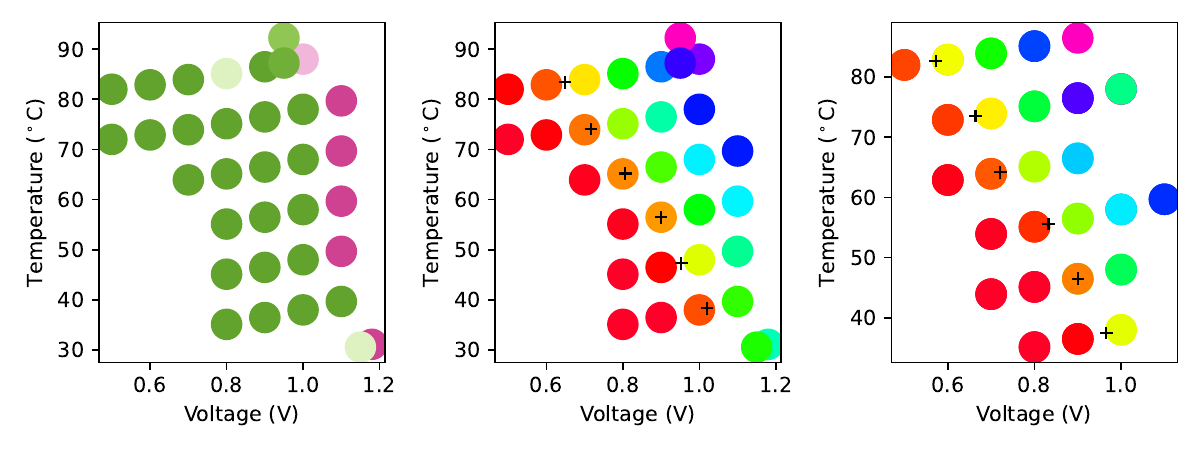}
    \caption{Phase diagrams as in the main text where the self heating of junctions due to the finite bias has been computed. This modifies the true temperature of the junctions during the tuning process resulting in a distortion of the phase diagram. The first panel shows the failure of junctions, the second panel the tuning of junctions at 1~kHz and the third panel the tuning of junctions at 103~kHz. }
    \label{fig:phase_diag_self_heat}
\end{figure*}
In the tuning process a current flows across the junction. Approximating the junction to a linear resistor gives $I_{\rm Tuning}(t) \sim V_{\rm Tuning}\cos(\omega t)/R_{\rm Junction}(t)$ where $V_{\rm Tuning}\cos(\omega t)$ is the oscillating voltage applied by the lock-in and $R_{\rm Junction}(t)$ is the resistance of the junction, which changes across the process. 

Electrons which tunnel across the barrier will have higher potential than the surrounding bath and will lose that excess energy as heat. We can estimate the lengthscale over which this  process occurs as equating to the mean free path of an electron in the metal, typically $\sim~10$~nm, i.e. very close to the oxide barrier. The power of this heat source is given by $P = V^2_{\rm Tuning}\cos(\omega t)^2/R_{\rm Junction}(t)$. Ignoring the time-varying junction resistance and computing the time average of this value we find $P = V^2_{\rm Tuning}/2R_{\rm Junction}$. For typical junction resistances of $R_{\rm Junction} \sim 5$~k$\Omega$ and an example tuning voltage of 1~V this equates to a heating power, $P \sim$ 0.1~mW, not insubstantial in a nanoscale device. 

We can estimate the change in temperature arising from this heat source by computing the temperature change arising from a point heat source in an infinite solid. $\delta T(r) = P/4\pi kr$ where $k$ is the thermal conductivity of the material ($\sim$30~Wm$^{-1}$K$^{-1}$ for sapphire at room temperature~\cite{dobrovinskaya2009properties}). To get a simple estimate of the change in temperature in a $\sim$200~nm device we find the average temperature change integrating the above function in spherical coordinates from 0 to 100~nm. Our device-on-substrate system better resembles a semi-infinite solid (i.e. heat can only flow in negative z, rather than positive and negative z). To approximate this we simply multiply the temperature differential by 2 assuming that heat transfer to air is negligible. For a bias of 1~V following the above protocol we find a device temperature change of $\sim$8~K. 

For each point in the phase diagram, we can adjust the temperature value by re-computing the heating power for the applied voltage. We can then add the temperature change to the average ambient temperature of the process. This modifies the phase diagram as shown in Fig.~\ref{fig:phase_diag_self_heat}, deforming the phase diagram. 

It remains difficult to fit the depinning model as presented in the main text to these deformed phase diagrams. Accurately accounting for these types of effect may be important when accurately modeling creep in these systems.

\section{Estimating $\tau$}
We can estimate the value of $\tau$ required for a qualitative match between experimental data and our model. To do this we take the ratio the creep-boundary $V_\omega$ i.e. Eq.~2 main text with the added $\Lambda$ denominator. This gives $T_{\rm 103 Hz}/T_{\rm 1000 Hz}$ which by arranging Eq.~2 main text and neglecting terms which approximate to unity we approximate to $\Lambda_{\rm 1000 Hz}/\Lambda_{\rm 103 Hz} = \ln{(1000\times2\pi\tau)}/\ln{(103\times2\pi\tau)}$. 
Equating this to the ratio of the y-intercepts for the two lines fitted to the stable/creep boundary in Fig.~2~(d) main text (424.95 K/437.95 K) we can solve for $\tau\sim3\times10^{-36}$~s - an unphysically small number.  

\bibliography{bibliography}% Produces the bibliography via BibTeX.